\documentclass[twoside,twocolumn,9pt]{article}
\usepackage{extsizes}
\usepackage[super,sort&compress,comma]{natbib} 
\usepackage[version=3]{mhchem}
\usepackage[left=1.5cm, right=1.5cm, top=1.785cm, bottom=2.0cm]{geometry}
\usepackage{balance}
\usepackage{mathptmx}
\usepackage{sectsty}
\usepackage{graphicx} 
\usepackage{lastpage}
\usepackage[format=plain,justification=justified,singlelinecheck=false,font={stretch=1.125,small,sf},labelfont=bf,labelsep=space]{caption}
\usepackage{float}
\usepackage{fancyhdr}
\usepackage{fnpos}
\usepackage[english]{babel}
\addto{\captionsenglish}{%
  
}
\usepackage{array}
\usepackage{droidsans}
\usepackage{charter}
\usepackage[T1]{fontenc}
\usepackage[usenames,dvipsnames]{xcolor}
\usepackage{setspace}
\usepackage[compact]{titlesec}
\usepackage{hyperref}
\usepackage{soul}

\usepackage{epstopdf}%This line makes .eps figures into .pdf - please comment out if not required.

\definecolor{cream}{RGB}{222,217,201}

\begin{document}

\pagestyle{fancy}
\thispagestyle{plain}
\fancypagestyle{plain}{
%%%HEADER%%%
\renewcommand{\headrulewidth}{0pt}
}
%%%END OF HEADER%%%

%%%PAGE SETUP - Please do not change any commands within this section%%%
\makeFNbottom
\makeatletter
\renewcommand\LARGE{\@setfontsize\LARGE{15pt}{17}}
\renewcommand\Large{\@setfontsize\Large{12pt}{14}}
\renewcommand\large{\@setfontsize\large{10pt}{12}}
\renewcommand\footnotesize{\@setfontsize\footnotesize{7pt}{10}}
\makeatother

\renewcommand{\thefootnote}{\fnsymbol{footnote}}
\renewcommand\footnoterule{\vspace*{1pt}% 
\color{cream}\hrule width 3.5in height 0.4pt \color{black}\vspace*{5pt}} 
\setcounter{secnumdepth}{5}

\makeatletter 
\renewcommand\@biblabel[1]{#1}            
\renewcommand\@makefntext[1]% 
{\noindent\makebox[0pt][r]{\@thefnmark\,}#1}
\makeatother 
\renewcommand{\figurename}{\small{Fig.}~}
\sectionfont{\sffamily\Large}
\subsectionfont{\normalsize}
\subsubsectionfont{\bf}
\setstretch{1.125} %In particular, please do not alter this line.
\setlength{\skip\footins}{0.8cm}
\setlength{\footnotesep}{0.25cm}
\setlength{\jot}{10pt}
\titlespacing*{\section}{0pt}{4pt}{4pt}
\titlespacing*{\subsection}{0pt}{15pt}{1pt}
%%%END OF PAGE SETUP%%%

%%%FOOTER%%%
\fancyfoot{}
\fancyfoot[LO,RE]{\vspace{-7.1pt}\includegraphics[height=9pt]{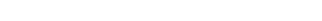}}
\fancyfoot[CO]{\vspace{-7.1pt}\hspace{11.9cm}\includegraphics{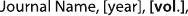}}
\fancyfoot[CE]{\vspace{-7.2pt}\hspace{-13.2cm}\includegraphics{head_foot/RF}}
\fancyfoot[RO]{\footnotesize{\sffamily{1--\pageref{LastPage} ~\textbar  \hspace{2pt}\thepage}}}
\fancyfoot[LE]{\footnotesize{\sffamily{\thepage~\textbar\hspace{4.65cm} 1--\pageref{LastPage}}}}
\fancyhead{}
\renewcommand{\headrulewidth}{0pt} 
\renewcommand{\footrulewidth}{0pt}
\setlength{\arrayrulewidth}{1pt}
\setlength{\columnsep}{6.5mm}
\setlength\bibsep{1pt}
%%%END OF FOOTER%%%
\newcommand{\erik}[1]{{\color{red}\small \bf EDH: \it #1 }}
\newcommand{\erikintro}[1]{{\color{red}\small \bf EDH suggests to introduce: [ {\color{gray} #1 } ] }}
\newcommand{\erikchange}[2]{{\color{red}\small \bf EDH suggests to change: [ {\color{gray} #1 } ] to [ {\color{blue} #2 } ] }}

%%%FIGURE SETUP - please do not change any commands within this section%%%
\makeatletter 
\newlength{\figrulesep} 
\setlength{\figrulesep}{0.5\textfloatsep} 

\newcommand{\topfigrule}{\vspace*{-1pt}% 
\noindent{\color{cream}\rule[-\figrulesep]{\columnwidth}{1.5pt}} }

\newcommand{\botfigrule}{\vspace*{-2pt}% 
\noindent{\color{cream}\rule[\figrulesep]{\columnwidth}{1.5pt}} }

\newcommand{\dblfigrule}{\vspace*{-1pt}% 
\noindent{\color{cream}\rule[-\figrulesep]{\textwidth}{1.5pt}} }

\makeatother
%%%END OF FIGURE SETUP%%%

%%%TITLE, AUTHORS AND ABSTRACT%%%
\twocolumn[
  \begin{@twocolumnfalse}
{\includegraphics[height=30pt]{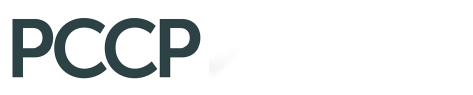}\hfill\raisebox{0pt}[0pt][0pt]{\includegraphics[height=55pt]{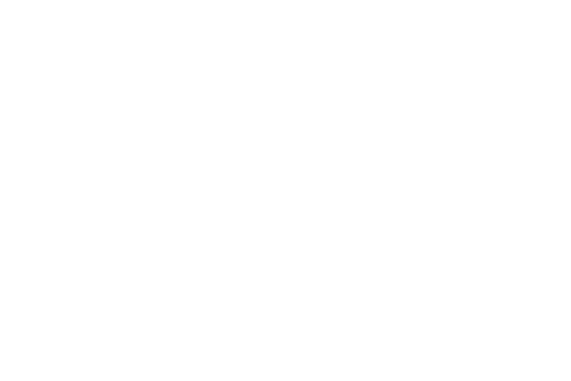}}\\[1ex]
\includegraphics[width=18.5cm]{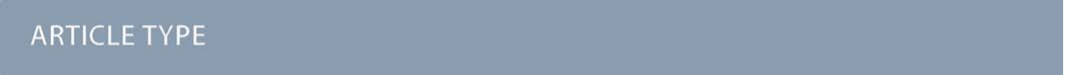}}\par
\vspace{1em}
\sffamily
\begin{tabular}{m{4.5cm} p{13.5cm} }

\includegraphics{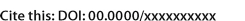} & \noindent\LARGE{\textbf{Perspective: Multi-configurational methods in bio-inorganic chemistry}} \\%Article title goes here instead of the text "This is the title"
\vspace{0.3cm} & \vspace{0.3cm} \\

 & \noindent\large{Frederik K. Jørgensen,\textit{$^{a}$} Micka{\"e}l G. Delcey,\textit{$^{b}$} and Erik D. Hedegård$^{\ast}$\textit{$^{,a,b}$}} \\%Author names go here instead of "Full name", etc.

\includegraphics{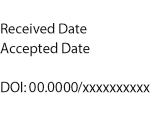} & \noindent\normalsize{
%Transition metals are an integral \textcolor{red}{part} of many bio-molecules and are responsible for \textcolor{red}{the function of many enzymes}\st{how many proteins function}. To ensure a physically sound description of the electronic structure of a metal active site in a protein, multiconfigurational wave functions are in principle  optimal. Yet, the computational cost of optimizing the multiconfigurational wave function has often prevented their use. However, in recent years, several developments making the optimization of multiconfigurational wave functions more efficient have been seen. In this perspective, we discuss some of these developments and how they have changed the use of multiconfigurational methods in bio-inorganic chemistry.
Transition metal ions play crucial roles in the structure and function of numerous proteins, contributing to essential biological processes such as catalysis, electron transfer, and oxygen binding. However, accurately modeling the electronic structure and properties of metalloproteins  poses significant challenges due to the complex nature of their electronic configurations and strong correlation effects. Multiconfigurational quantum chemistry methods are, in principle, the most appropriate tools for addressing these challenges, offering the capability to capture the inherent multi-reference character and strong electron correlation present  in bio-inorganic systems. Yet their computational cost has long hindered wider adoption, making methods such as Density Functional Theory (DFT) the method of choice. However, advancements over the past decade have substantially alleviated this limitation, rendering multiconfigurational quantum chemistry methods more accessible and applicable to a wider range of bio-inorganic systems. In this perspective, we discuss some of these developments and how they have already been used to answer some of the most important questions in bio-inorganic chemistry. We also comment on ongoing developments in the field and how the future of the field may evolve.
}\\

\end{tabular}

 \end{@twocolumnfalse} \vspace{0.6cm}

  ]
%%%END OF TITLE, AUTHORS AND ABSTRACT%%%

%%%FONT SETUP - please do not change any commands within this section
\renewcommand*\rmdefault{bch}\normalfont\upshape
\rmfamily
\section*{}
\vspace{-1cm}

%%%FOOTNOTES%%%

\footnotetext{\textit{$^{a}$~Department of Physics, Chemistry, and Pharmacy, University of Southern Denmark, Campusvej 55, 5230 Odense, Denmark; Tel: +45 20 44 69 23; E-mail: erdh@sdu.dk}}
\footnotetext{\textit{$^{b}$~Department of Chemistry, Lund University, Naturvetarvägen 14, 221 00 Lund, Sweden.}}

%Please use \dag to cite the ESI in the main text of the article.
%If you article does not have ESI please remove the the \dag symbol from the title and the footnotetext below.
%\footnotetext{\dag~Electronic Supplementary Information (ESI) available: [details of any supplementary information available should be included here]. See DOI: 10.1039/cXCP00000x/}
%additional addresses can be cited as above using the lower-case letters, c, d, e... If all authors are from the same address, no letter is required

%\footnotetext{\ddag~Additional footnotes to the title and authors can be included \textit{e.g.}\ `Present address:' or `These authors contributed equally to this work' as above using the symbols: \ddag, \textsection, and \P. Please place the appropriate symbol next to the author's name and include a \texttt{\textbackslash footnotetext} entry in the the correct place in the list.}

%%%END OF FOOTNOTES%%%

%%%MAIN TEXT%%%%
\section*{Introduction}

Large parts of modern medical and industrial research utilize bio-molecules in the form of proteins and enzymes.\cite{bell2021} Predicting the properties of these bio-molecules and their interactions with either substrates or external stimuli has been termed a "holy grail" in chemical modeling.\cite{houk2017} A predictive model can uncover relations between molecular structure, function, and mechanism, which are key in rational design. 

Today it is possible to predict properties and interactions on an atomic level through quantum mechanical (QM) and classical models. This impressive performance was for long restricted to simple systems such as organic molecules the in gas phase.\cite{szabo2015,schreiner2015} However, in recent years increasingly complex bio-molecular systems have been modelled. Simulation techniques have accordingly become an integral part of drug discovery, occasionally even driving the development\cite{sadybekov2023}. This is achieved by increasingly sophisticated computer algorithms, combined with the efficiency of modern computer chips. Combined with machine learning and deep learning methods, these techniques are expected to be the backbone of drug discovery in the coming years. \cite{pandey2022} Despite this impressive development, certain bio-molecular systems still pose large challenges for theoretical methods. In particular, we are faced with severe issues for transition-metal-containing systems.\cite{pierloot2011,riccardi2018,grimme2018,sabe2021} This is dismaying seeing that one-third of all enzymes contain transition metals\cite{holm1996}, mediating vital processes, both in nature and industry. 

% Also mention severe issues with metal in MM?
The issue with describing transition metals in bio-molecular systems is that a sufficiently accurate description requires a QM method that can handle more than one dominant electronic configuration in a balanced way. Cases where multiple electron configurations are important are known as a \textit{strong} or \textit{static} correlation cases. In molecules with little static correlation, Hartree-Fock (HF) is considered \textit{qualitatively} correct and second-order Møller-Plesset (MP2) or the popular Kohn-Sham Density Functional Theory (DFT) are expected to give \textit{quantitatively} correct results. These two methods are accurate  since they capture the main part of the so-called \textit{dynamical} correlation well. In case even more accurate methods are required for dynamical correlation, the hierarchy of coupled cluster (CC) methods\cite{bartless2024} can be applied. Yet, DFT methods are much more efficient, making them more popular. However, for cases with large static correlation, MP2, DFT and CC methods break down\cite{neese2007,khedkar2021} due to their inherent approximation that the electronic state is described well by a single, dominant electronic configuration.

Wave functions tailored to capture static correlation are denoted as multiconfigurational. Developing these wave function methods is an active field of research\cite{ghosh2018,park2020}, but their large computational cost remains an obstacle. One of the challenges is to make the methods efficient enough  to describe the transition metal active site of proteins. One requirement is that the methods can include a sufficiently large amount of configurations. In recent years, several new developments have focused on restricting the number of configurations, while devising a physically motivated algorithm to include \textit{the right} configurations. Including the right configurations is a major challenge for the traditional multiconfigurational wave functions. Yet, with these recent advancements, calculations with multiconfigurational wave functions on transition metal clusters such as the \ce{Mn4CaO5} cluster from photosystem II have been achieved -- something that would have been completely unthinkable some years before.\cite{kurashige2013} Another impressive landmark was when the iron-sulfur clusters present in numerous proteins (with up to eight spin-coupled iron atoms) were described with a multiconfigurational wave function\cite{sharma2014b,li2019} showing that the clusters indeed are strongly correlated. 

While calculations of metal clusters are significant achievements, the above-mentioned theoretical work only includes dynamical correlation to a limited extent. Efficient treatment of dynamical correlation is another challenge for traditional multiconfigurational methods. Another (largely overlooked) challenge is the development of efficient algorithms to treat a large number of atoms (i.e. large number of basis functions). This is critical to ensure that the methods can applied beyond model systems. 

In this perspective, we discuss the current status of multiconfigurational methods focusing on bio-inorganic systems. Recent perspectives have focused on multiconfigurational methods more broadly for (molecular) inorganic compounds\cite{khedkar2021}, or multiconfigurational and relativistic methods for inorganic compounds in chemotherapy\cite{hedegaard2022}. Therefore, our focus here is mainly on advancements for protein metal active sites. Unlike the chemotherapeutic compounds that often contain heavy metals such as platinum, the vast majority of metalloproteins harbors metals from the 3d row, where relativistic effects are expected to be benign. Accordingly, we mainly focus on non-relativistic methods, with one notable exception: theoretical treatment of certain core spectroscopies is not possible without the inclusion of relativistic effects. Thus, one section briefly covers the differences and challenges introduced when describing core spectroscopy  with a multiconfigurational wave function. 

We first summarize the recent developments in the methods, both improvements towards a large number of configurations and new methods for the efficient description of dynamical correlation, even allowing the treatment of  full proteins with multiconfigurational methods. A brief section then discuss how relativistic effects can be included. Finally, we illustrate the improved capabilities with selected case studies with proteins, containing metal clusters as well as active sites with a single metal atom.  %e will here discuss a different example, namely an iron-nickel hydrogenase, since these have been treated with a variety of multiconfigurational methods,  allowing some comparison of the different flavors of the methods. We also discuss cases where the reaction center architecture demands many configurations to describe a facile interaction between the ligand and a single metal center.

%\section*{Correlation in metal complexes}
%Metal complexes are some of the most challenging molecules for quantum chemistry. One of the main reason for this is that electron correlation has a particularly large influence on their properties, and this correlation may be both in the form of dynamical and strong/static correlation.

%Strong correlation is often exemplified using bond dissociation. As a covalent bond is stretched towards homolytic dissociation, the energy gap between the bonding and antibonding orbitals shrink and the wavefunction becomes more and more multideterminantal. However, a similar effect can happen even at equilibrium geometries, and this is particularly common for transition metals.
%\textcolor{red}{Mickael}

\section*{The complete active space wave function}

The QM methods discussed in this perspective generally seek a wave function, $|\Psi\rangle$, that is a solution to the time-independent Schrödinger equation 
\begin{equation}
	\hat{H}\vert \Psi\rangle  = E \vert \Psi\rangle . 
	\label{eq:schrodinger}
\end{equation}
The wave function can be constructed in various ways from a set of $M$ orthonormal molecular spin orbitals (MOs) within a Fock subspace, $F(M,N)$, with a fixed number of $N$ electrons.\cite{helgaker2004} The most general solution is obtained as the linear combination of all possible ways of distributing the $N$ electrons within the $M$ MOs.  We write this linear combination as  
\begin{equation}\label{full-ci}
	|\Psi\rangle=\sum_{i}c_{i}|\psi_{i}\rangle    ,
\end{equation}
and this is known as the full configuration interaction (Full-CI) state. Here, $|\psi_{i}\rangle $ represents a given electronic configuration (e.g. given by a Slater determinant) with weight $c_i$. The parametrization of multiconfigurational wave functions is conveniently expressed in a second-quantization formulation.\cite{helgaker2004} In this formulation, the (non-relativistic) Hamiltonian in Eq.~\eqref{eq:schrodinger} is given
\begin{equation}
	\hat{H} = \sum_{pq} h_{pq} \hat{E}_{pq} + \frac{1}{2}\sum_{pqrs} g_{pqrs}\left( \hat{E}_{pq} \hat{E}_{rs} - \delta_{qr} \hat{E}_{ps} \right) + V_{\text{nn}}
	\label{hamilton} , 
\end{equation}
where $h_{pq}$ and $g_{pqrs}$ are the standard one- and two-electron integrals  and $\hat{E}_{pq}=    \hat{a}_{p\alpha}^\dagger\hat{a}_{q\alpha} + \hat{a}_{p\beta}^\dagger\hat{a}_{q\beta}$ is defined by means of  creation ($\hat{a}^{\dagger}_{p\sigma}$) and annihilation ($\hat{a}_{q\sigma}$) operators, respectively. The indices $p,q,r,s$ denote general (spatial) orbitals.  The energy from Eq.~\eqref{eq:schrodinger} can be obtained as an expectation value over $\vert \Psi\rangle$, i.e., by means one- and two-electron reduced density matrices 
\begin{align}
	D_{pq} &= \langle\Psi\vert \hat{E}_{pq} \vert \Psi\rangle \label{eq:one-elec-dens} \\
	P_{pqrs} &= \langle\Psi\vert \hat{E}_{pq} \hat{E}_{rs} - \delta_{qr} \hat{E}_{ps}  \vert \Psi\rangle. \label{eq:two-elec-dens}
\end{align}

\begin{figure}[htb!]
	\centering
	\includegraphics[width=0.5\textwidth]{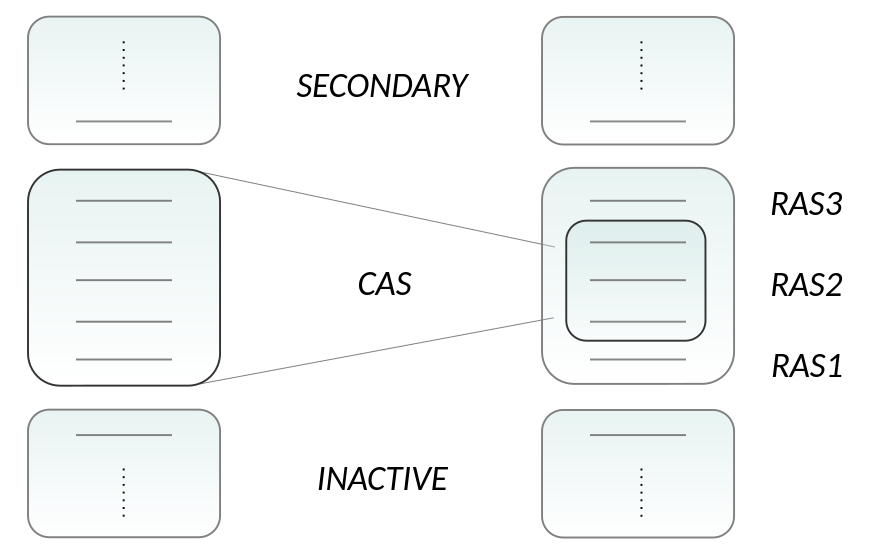}
	\caption{Division of the orbital space into inactive, active, and secondary orbitals. An FCI calculation is performed within the active orbital space.}
	\label{fig:CAS-RAS-fig}
\end{figure}

Whilst $|\Psi\rangle$ may be the true electronic state (within the given set of MOs), the number of configurations grows factorially with respect to the system size and quickly becomes computationally unfeasible, even for small  systems. Most modern multi-configurational (MC) methods handle this problem by partitioning the full orbital space into subspaces with the most popular partitioning scheme being the complete active space (CAS) scheme.\cite{roos1980c}
In the CAS method, the full orbital space is partitioned into \textit{three} subspaces usually denoted as the \textit{inactive}, \textit{active}, and \textit{virtual} orbital spaces, each containing a fixed set of MOs for a given molecular system (see Figure \ref{fig:CAS-RAS-fig}). For the inactive subspace, occupation numbers for all orbitals are two, corresponding to doubly occupied orbitals. For the virtual subspace, all occupation numbers are zero (corresponding to empty orbitals), whereas the subset of orbitals within the active space are represented by a Full-CI (FCI) wave function, resulting in  occupation numbers between zero and two. Only performing an FCI within this active space greatly reduces the computational cost of the \textit{ansatz} in Eq.~\eqref{full-ci}. 
The remaining (inactive or virtual) orbitals are either kept fixed or variationally optimized in a self-consistent field (SCF) procedure so they can adapt (relax) to the final, optimized form of the CAS wave function. Generally, keeping the inactive and virtual orbitals fixed is denoted CAS-CI, whereas using a CAS with SCF optimized orbitals is denoted CASSCF.
  
The CAS wave function thus takes the form of Eq. \eqref{full-ci} within a constrained orbital space spanned by the set of active orbitals $\{\phi\}$,
\begin{equation}\label{cas_wf}
	|\mathrm{CAS}\rangle=\sum_{i}^{L}c_{i}|\psi_{i}\rangle  
\end{equation}
where $ \psi_i $ again is a determinant (or configuration state function) and $L$ denotes the total number of possible determinants (or configuration state functions). % contributing to the electronic state, determined by a given symmetry, spin multiplicity, and number of electrons within the active space.

With the increase in computational power and the development of highly efficient, parallel algorithms, the size limitations of modern CAS spaces have reached upwards of 20 electrons in 20 active orbitals.\cite{vogiatzis2017pushing, delcey23-multipsi} 
Although this enables over 100 billion possible configurations, the computational simplification of the \textit{ansatz} within Eq. \eqref{cas_wf} as compared with the (often unfeasible) ansatz of Eq. \eqref{full-ci} comes at a price, since the quality of the CAS calculation will depend on the chosen orbitals within the active subspace. Selecting  the ``correct'' set of orbitals, or even determining what constitutes "correct", represents a major challenge within the CAS scheme, and will inevitably lead to some user bias. Guidelines based on chemical intuition\cite{veryazov2011} or semi-automatic schemes based on non-biased, analytical metrics\cite{stein2016} have been proposed to guide users towards a suitable, non-biased active space of orbitals.

Even with automatic procedures for the selection of an active space, a  drawback of CAS-based methods is that many protein metal active sites require active spaces beyond what is currently possible. An interesting up--coming approach is to simulate the CI problem on a quantum device, but this technology still requires to mature before the potential within bio-inorganic chemistry can be properly assessed.\cite{lee2023} We therefore  focus on multiconfigurational methods relying on classical computers. For these methods, the issues related to the size of the required active space have been addressed by approximating the FCI by imposing restrictions on which configurations that are included; this again allows extending the active space.
We discuss these methods in the next subsection.
\begin{figure}[htb!]
	\centering
	\includegraphics[width=0.4\textwidth]{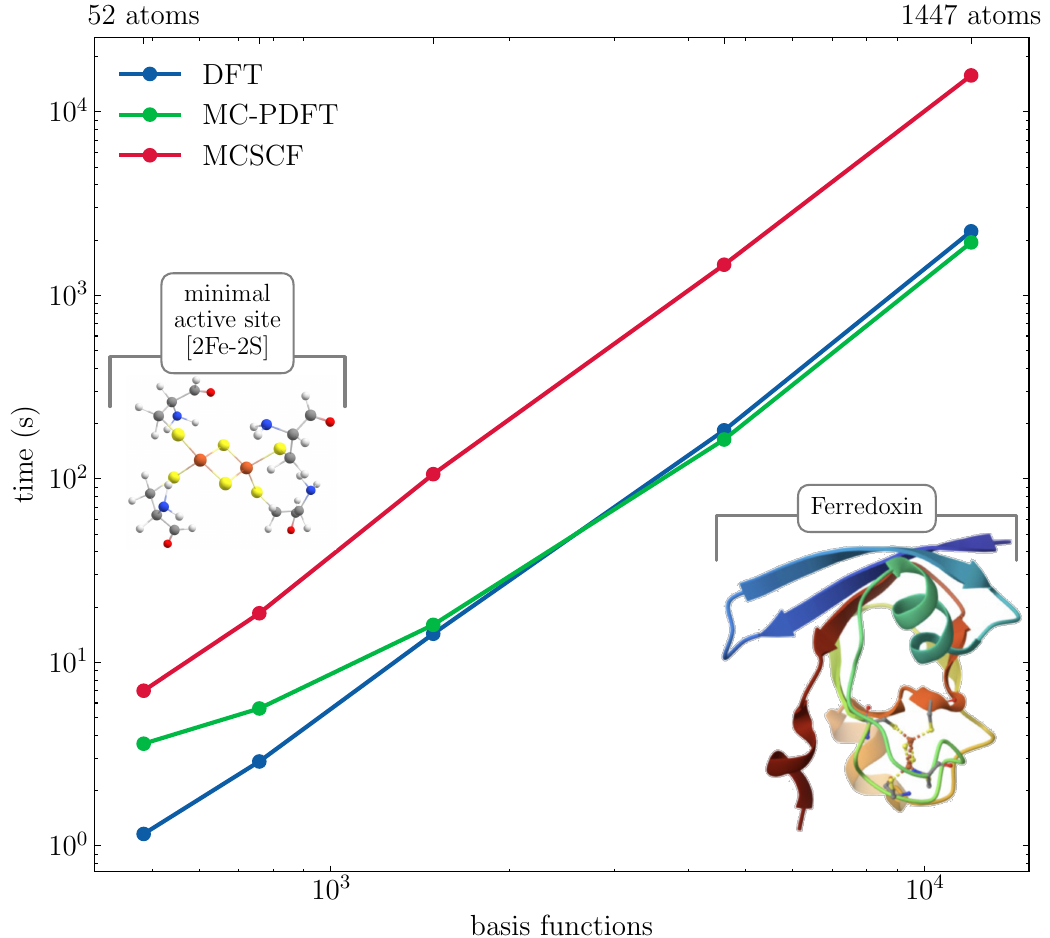}
	\caption{Computational time per iteration for DFT, CASSCF and CAS-PDFT using increasing model sizes of the ferredoxin protein. The calculations were performed using VeloxChem/MultiPsi\cite{rinkevicius2020veloxchem, delcey23-multipsi} on 8 nodes consisting each of two 16 cores Intel Xeon Gold 6130 CPUs, for a total of 256 cores. The last point with 1447 atoms corresponds to the entire protein. Figure taken from reference \citenum{scott24-varpdft}.}
	\label{fig:pdft_scaling}
\end{figure}

For metal active sites that do not necessarily warrant the use of a large active space, the wave function optimization may still  involve a vast number of inactive orbitals. For a given active space, a CASSCF calculation should have the same computational scaling as Hartree-Fock, yet, development of efficient codes have somewhat lagged behind. This is an often overlooked issue with CASSCF wave functions. The situation has improved significantly the last years with  efficient implementations using the resolution of the identity or the related Cholesky Decomposition,\cite{nottoli21-cd-casscf, helmichparis22-casscf} as well as parallel or even GPU implementations.\cite{hohenstein15_GPU-casscf} We here highlight the program {\sc MultiPsi}\cite{delcey23-multipsi} which has recently been developed to handle large-scale multiconfigurational wave functions. Treating systems with over 500 atoms is routine in {\sc MultiPsi} and molecules upwards of a thousand atoms can be simulated. We show an example in Figure \ref{fig:pdft_scaling} where an entire metalloprotein (ferredoxin, 1447 atoms) has been optimized with a CASSCF wave function.\cite{scott24-varpdft} These developments will hopefully allow us to move from the small model systems employed in refs \citenum{kurashige2013,sharma2014b,li2019} to more realistic models of the active sites (and eventually entire proteins).

\section*{Approximate complete active space methods}

The simplest of the approximate FCI methods is the restricted active space (RAS), cf.~Figure \ref{fig:CAS-RAS-fig}, where the FCI only is performed within RAS2, whereas RAS1 and RAS3 have a (user-defined) limit on which configurations that can contribute.\cite{malmqvist1990,malmqvist2008} An often employed setting is allowing configurations corresponding to two holes in the RAS1 and two electrons in RAS3. The idea has also been generalized to an arbitrary number of active spaces with different restriction levels.\cite{ma2011}  The RASSCF method has been around for some time, so we focus here instead on newer approximate FCI solvers, namely the FCI quantum monte-carlo (FCI-QMC) and density matrix renormalization group (DMRG) methods. In recent years, FCI-QMC and DMRG have made calculations tractable that previously were prohibitively expensive. The main advantage of FCI-QMC and DMRG is that they perform an approximate FCI, where the errors are easier to control than in RASSCF.

Starting with DMRG, this method was originally developed for applications in solid-state physics. Yet, it has in recent years demonstrated impressive performance for static-correlation problems in quantum chemistry.\cite{marti2011,chan2011,wouters2014rev,knecht2016} The initial developments were done by White\cite{white1992b,white1993} as a further development of Wilson’s numerical renormalization group.  In DMRG terminology, each spatial orbital within the active space defines a \textit{site} with four possible one-electron states, $|n_i \rangle = \{ |  vac\rangle, |\alpha\rangle, |\beta\rangle, | \alpha\beta\rangle \}_i $.  Thus, the CAS wave function is defined as an occupation number vector,  $\vert\Psi \rangle=|n_1, n_2,\ldots, n_L\rangle $. We can then reformulate the FCI \textit{ansatz} in Eq.~\eqref{cas_wf} as
\begin{equation}\label{cas_wf_tensor}
|\mathrm{CAS}\rangle=\sum_{n_1,n_2,\ldots,n_L }C_{n_1 n_2\ldots n_L} |n_1 , n_2,\ldots, n_L\rangle  ,
\end{equation}
which is possible for any state in a Hilbert space spanned by $L$ spatial orbitals. In Eq.~\eqref{cas_wf_tensor}, the CI  coefficients, $c_i$, are written as a coefficient tensor,   $C_{n_1 n_2\ldots n_L} $ . This tensor can be formulated as a matrix product so that the wave function in Eq.~\eqref{cas_wf_tensor}  becomes 
\begin{equation}
	\label{cas_wf_tensor_mps}
	|\mathrm{CAS}\rangle=\sum_{n_1,n_2,\ldots,n_L } M^{n_1}M^{n_2}\ldots M^{n_L} |n_1 , n_2,\ldots, n_L\rangle  .
\end{equation} 
The state in Eq.~\eqref{cas_wf_tensor_mps} is a so-called matrix product state (MPS), where the CI coefficients are now encoded in the matrices $\{M^{n_i}\}$ instead of the tensor in Eq.~\eqref{cas_wf_tensor}. The DMRG algorithm optimizes these matrices (and by analogy the CI coefficients) variationally through successive \textit{sweeps} from left to right (and vice versa) until convergence is reached. In fact, the variational optimization of the DMRG \textit{ansatz} and variational optimization of an MPS wave function \textit{ansatz} was first shown to be equivalent after the original introduction of DMRG\cite{ostlund1995,rommer1997,verstraete2004,knecht2016}, where the DMRG algorithm was not as directly linked to an MPS as in Eq.~\eqref{cas_wf_tensor_mps}.

The dimensions of the matrices $\{M^{n_i}\}$  in Eq.~\eqref{cas_wf_tensor_mps}  are generally $m_{i-1}\times m_i$ (note that $M^{n_1}$ and  $M^{n_L}$ are restricted to the dimensions $1\times m_1$ and $m_{L-1}\times 1$ to ensure that the matrix product becomes a scalar). The efficiency of DMRG comes from choosing an upper limit, $m$, for the matrix dimension of $\{M^{n_i}\}$.  The upper limit is called \textit{number of renormalized states} and is typically (many) orders of magnitude smaller than solving the FCI problem. Without restrictions on $m$, the FCI solution within the chosen CAS is recovered, but all efficiency gains are lost. The FCI result can be approached systematically by increasing $m$. The states that are kept during the sweeps are chosen based on their weight in the density matrix. Notably, other selection criteria were employed in earlier renormalization group methods, but employing the density matrix significantly improves the performance.

An alternative method to obtain approximate results for large CI expansions is the full configuration interaction quantum Monte-Carlo (FCI-QMC).\cite{booth09-fci-qmc} Like standard quantum Monte-Carlo (QMC), this method is based on a stochastic imaginary-time propagation of the Schrödinger equation, but unlike standard QMC, this is done in the space of Slater Determinants. In practice, the method generates a set of walkers that sample the CI wavefunction, approximating the CI coefficients by the average population (and sign) of each walker. Similar to the use of the density matrix during the sweeps in DMRG, the spawning criteria ensures that the algorithm samples the most significant Slater Determinants. In this sense, the method has similarities with selected CI techniques\cite{huron73-selectedci, harrison91-selci}, making use of the sparsity of the CI vectors.\cite{ivanic01-ci-deadwood} While FCI-QMC is often done in the basis of Slater-Determinant, spin-pure implementations using purification techniques\cite{Weser22-spin-pure-fciqmc} or using a basis of configuration state functions\cite{dobrautz21-guga-fciqmc} have also been implemented.

The DMRG and FCI-QMC methods can be used as FCI solvers for CASSCF. For metal active sites, they are mostly used when very large active spaces are demanded, either due to a intricate couplings between the ligand and the d-electrons or if multiple metals are present. We will discuss examples in the case studies below. We will generally denote DMRG as DMRG-CI if the method does not include SCF orbital optimization. In case of FCI-QMC, the resulting CASSCF is often coined stochastic CASSCF.\cite{thomas15-stoch-mcscf,limanni2016}

\section*{Multiconfigurational perturbation theory}

While modern, approximate FCI solvers described in the previous section have pushed the limit of the active space sizes that can be treated, this still only covers a small fraction of the correlation in a typical bio-inorganic system. Therefore, we need an accurate treatment of the correlation outside of the active space. If the active space has been chosen correctly, this remaining part of the correlation will mainly be \textit{dynamical} correlation as mentioned in the introduction (although we note that the division of correlation into static/strong and dynamical parts is purely phenomenological).

Within wavefunction theory, a method to recover dynamical correlation that is \textit{relatively} affordable, is second-order multireference perturbation theory. Several variants have been suggested throughout the years, with the most popular being the complete active space perturbation theory (CASPT2)\cite{andersson1990, andersson1992} and the N-electron valence perturbation theory (NEVPT2)\cite{angeli2001,angeli2002-nevpt2}. The different multireference perturbation theories differ mainly in the choice of zeroth order Hamiltonian and how the interacting space is generated.

The standard variants of CASPT2 and NEVPT2 use a partially or fully internally contracted framework, meaning that the interacting space is generated by applying excitation operators directly on the CASSCF (or RASSCF)  wavefunction, and not individually on every configuration in the reference space. This generates a significantly more compact ansatz, with only a moderately larger interacting space than a single-reference perturbation theory, and is thus key to an efficient method. The downside is that the resulting perturber wavefunctions are significantly more complicated and computing the required matrix elements necessitates high-order density matrices in the active space, typically third and fourth order. For large active spaces, the computation and storing of these density matrices can become a bottleneck, which partially negates the potential of the approximate active space solvers mentioned in the previous section. Cumulant approximation of the fourth order density matrix has been attempted,\cite{zgid09-cumulant, kurashige14-cumulant} with mixed results, especially in bio-inorganic systems.\cite{phung16-cumulant} However, recent works have demonstrated that the fourth order density matrix is not required and that both CASPT2 and NEVPT2 can be computed without any approximation using instead an intermediate resembling the third order density matrix.\cite{kollmar21-4DM}

When it comes to the choice of the zeroth order Hamiltonian, $H_0$, most are designed to be equivalent to M{\o}ller-Plesset for a single reference, and thus the difference lies in how the active orbitals are described. In CASPT2, a single-electron Fock-like operator $\hat{F}$ is chosen as $H_0$, see Eq.~\ref{eq:caspt2_fock}.
\begin{equation}
    \hat{F}_{pq} = \hat{a}_p [\hat{H}, \hat{a}_q^\dag] - \hat{a}_p^\dag [\hat{H}, \hat{a}_q] .
    \label{eq:caspt2_fock}
\end{equation}
While this makes the approach computationally efficient, it has several drawbacks. First, while the orbital energy of occupied orbitals represents an approximation of the ionization potential (IP) and that of virtual orbitals represents the electron affinity (EA), the CASPT2 formulation generates active orbital energies that are a sort of weighted average of the two, regardless of whether the correlating function corresponds to an excitation into or out of this orbital. This has shown to lead to systematic errors and led to the development of the ionization potential electron affinity (IPEA) shift.\cite{ghigo2004-ipea} While initial results appeared promising, the universality of the shift was later questioned. Recent results in organic spectroscopy suggested removing the shift altogether.\cite{zobel17-ipea} For transition metals, it was instead often found that a higher shift improved the results, especially concerning spin-state energetics, which was found to be due to a better description of 3s and 3p correlation at the expense of a worse description of the valence.\cite{pierloot17-3s3p} A second major issue with the CASPT2 definition of $H_0$ is the problem of intruder states. This arises whenever the (approximated) energy of a perturber state approaches that of the reference state, creating a singularity. This is typically addressed by using an imaginary level shift\cite{forsberg97-imls} or more recently through a regularization technique.\cite{battaglia22-regularized}

The problem of the IPEA and intruder states is to a large extent due to the one-electron nature of the chosen $H_0$ (Eq.~\ref{eq:caspt2_fock}). This has motivated the development of a scheme using a two-electron operator, but keeping the computational simplicity of CASPT2, leading to the NEVPT2 approach. More specifically, NEVPT2 uses a Dyall Hamiltonian.\cite{dyall95} However by using increasing levels of contraction of the interacting space, we still obtain an efficient implementation, and in its strongly contracted form, NEVPT2 is even more efficient than CASPT2 by not being iterative.\cite{angeli2002-nevpt2} NEVPT2 is also strictly size-consistent, unlike CASPT2 which is only approximately so. Despite these clear advantages, recent benchmarks have shown that CASPT2 performs typically better on both excitation energies\cite{sarkar22-caspt2_vs_nevpt2} and metal complexes\cite{pierloot17-3s3p}.

%Should I discuss MS/XMS, etc...

\section*{Multiconfigurational density functional theory}

Most multireference methods can be thought of as generalizations of a single-reference counterpart. In this sense, CASSCF corresponds to HF and CASPT2 to MP2. Considering the dominance of DFT among single reference methods, it is no surprise that a multireference variant of DFT has been long sought after. This field is still under very active developments, with several formulations currently competing.\cite{hedegaard20-mcdft,ghosh2018} However, all of these multiconfigurational DFT variants share the same potential of treating both strong/static and dynamical correlation at a low cost.

The earliest approach can be traced back to the 1970s, where Lie and Clementi suggested to simply add a correlation functional to an energy obtained from a multiconfigurational wave function.\cite{lie74} The motivation was that the DFT correlation functional only describes the dynamical correlation while the wave function only treats the strong correlation. However, the latter assumption is poor as any multiconfugrational wave function inevitably includes parts of the dynamical correlation within the active space, leading to what was coined the double-counting problem.\cite{cremer2001-density} Since then, various approaches have been suggested to avoid double counting.

A natural idea is to estimate and remove the DFT contribution from the active space. This requires adding one (or several) new parameter(s) to the functional as a measure of the active space. One such idea is the CAS-DFT method, where a scaling factor is added to the functional. The scaling factor depends on the inactive density and/or on what the authors call "the reference density", which is the density if all active orbitals were occupied.\cite{savin88-dft_ci,miehlich97-dft_ci,grafenstein00-casdft} Another idea is to directly measure the amount of correlation included using the on-top pair-density, $\Pi(r)$. The on-top pair density can be extracted from a multiconfigurational wave function by using the overlaps of spatial orbitals, $\Omega_{pq}(r)$, and the two-electron reduced density matrix in Eq.\eqref{eq:two-elec-dens}   obtained with a CAS(SCF) wave function
\begin{equation}
\Pi(r) = \sum_{pqrs}\Omega_{pq}(r) \Omega_{rs} (r) P_{pqrs}.
\label{eq:ontop}
\end{equation}
The on-top pair density is related to the probability of two electrons being in the same point in space.\cite{gusarov04-mcdft} This concept is the key idea behind the CAS$\Pi$DFT method of the group of K. Pernal.\cite{gritsenko18-caspidft} A common difficulty in these approaches is to determine the resulting densities-dependent scaling factor.

A fundamentally different approach to solve the double-counting problem has been to completely forego the CASSCF energy and instead solve the strong correlation problem completely within a DFT framework. This idea traces back to the work of Moscard\'o \emph{et al} \cite{moscardo91_pdft} and later Becke \emph{et al}\cite{becke1995-ontop} who argued that for open-shells, the on-top pair-density may be a better functional argument than spin-densities, in particular in cases of bond breaking and multiplet splitting. Some authors even argued the success of current spin-density functional approximations was because they approximate the on-top pair-density and thus the exchange hole.\cite{perdew1995} This led in 2014 to the multiconfigurational pair-density functional theory method (MC-PDFT) which has since grown in popularity.\cite{manni2014, sharma21-mcpdft} In this approach, the energy is computed by a Kohn-Sham-like formula:
\begin{equation}
	E_\mathrm{pdft} = \langle \mathrm{CAS} | \hat{h} + \hat{J} | \mathrm{CAS} \rangle + \int E_\mathrm{xc}(\rho, \Pi) dr
	\label{eq:pdft}
\end{equation}
with $\hat{h}$ and $\hat{J}$ being the one-electron and Coulomb operator, respectively, and $E_\mathrm{xc}$ the PDFT functional which depends on the total density, $\rho(r)$, and the on-top pair-density, $\Pi(r)$. Extracting the on-top pair density from a CAS(SCF) wave function can be done according to Eq.~\eqref{eq:ontop}, while the density can be extracted as
\begin{equation}
	\rho(r) = \sum_{pq}\Omega_{pq}(r)  D_{pq}. 
	\label{eq:dens}
\end{equation}
Here, $  D_{pq}$ is one-electron reduced density matrix in Eq.~\eqref{eq:one-elec-dens}, obtained with a CAS(SCF) type wave function. The PDFT functional can be derived from scratch or it can be adapted from standard spin-density functionals by using a simple "translation formula":
 
\begin{equation}
    \label{eq:pdm_translation_full}
    \rho_{\alpha/\beta} = \frac{\rho \pm \sqrt{\rho^2 - 2 \Pi}}{2}
\end{equation}

Note that this formula needs to be adapted to address some issues when going beyond the local density approximation\cite{carlson2015-ft_pdft} as well as for some systems such as low-spin open-shells or excited states.\cite{rodrigues23_complex} Traditionally, the energy in Eq.~\eqref{eq:pdft} was computed as a non-variational correction after a wavefunction optimization at the CASSCF level, but recently a variationally optimized MC-PDFT implementation has also been derived.\cite{scott24-varpdft} In both cases, as no CASSCF energy is used, the result is formally free from double counting. The CASSCF energy can be added back to create a hybrid formulations but this requires re-introduction of scaling factors to avoid double counting.\cite{pandharkar20-hmc_pdft} In several benchmarks, MC-PDFT has been found to perform well, with results often on-par with CASPT2 despite its much lower computational cost.\cite{sharma21-mcpdft} The reduced cost comes from both the reduced scaling with respect to the number of inactive orbitals as well as the fact that only the density and on-top pair density are used for the functional in Eq.~\eqref{eq:pdft}. These quantities only require the one- and two-particle reduced density matrices (as can be seen from Eqs.~\ref{eq:dens} and \ref{eq:ontop}). Thus, the dynamical correlation is obtained without the need for higher-order reduced density matrices. For small to medium active spaces, the variational MC-PDFT can (efficiency-wise) be competitive with (unrestricted) DFT, making MC-PDFT calculations for more than thousand atoms possible. This was recently demonstrated using the implementation in {\sc MultiPsi} (see Figure \ref{fig:pdft_scaling}).

 A different solution to the double-counting problem is represented by the multiconfigurational short-range
DFT (MC--srDFT) method.\cite{savin1995,savinbook,toulouse2004long,fromager2007,hedegaard2015b} In the MC--srDFT method, the double-counting problem is avoided by splitting the electron-electron interaction into a long range and short range contribution using the error function,
\begin{equation}
    \frac{1}{r_{ij}}=\frac{\mathrm{erf}(\mu r_{ij})}{r_{ij}}-\frac{1-\mathrm{erf}(\mu r_{ij})}{r_{ij}},
    \label{eq:electron-rep-range}
\end{equation}
where $\mu$ is denoted the \textit{range-separation parameter}. This separation allows for a functional treatment of the short range electron-electron interactions whilst the long range part is treated by a CASSCF wave function.\cite{savin1995,savinbook,toulouse2004long}  The MC--srDFT energy can thus be written as
\begin{equation}
    E =\langle\Psi^{\mathrm{lr}}|\hat{H}^{\mathrm{lr}}|\Psi^{\mathrm{lr}}\rangle+E_{\mathrm{H}}^{\mathrm{sr}}\left[\rho\right]+E_{\mathrm{xc}}^{\mathrm{sr}}\left[\rho\right] , 
    \label{eq:energy-range-seperated}
\end{equation}
where the short-range Hartree functional $E_{\mathrm{H}}^{\mathrm{sr}}\left[\rho\right]$
and exchange-correlation,$E_{\mathrm{xc}}^{\mathrm{sr}}\left[\rho\right]$,
are complimentary to the long range expectation value in the sense
that $E =E^{\mathrm{lr}}+E^{\mathrm{sr}}$. The long-range Hamiltonian, $\hat{H}^{\mathrm{lr}}$, is identical to Eq.~\eqref{hamilton}, except that the two-electron repulsion integrals employ the long-range part of the electron repulsion operator in Eq.~\eqref{eq:electron-rep-range}. The electron density employed in the functionals can be obtained from a multiconfigurational wave function through the one-electron reduced density matrix according to Eq.~\eqref{eq:dens}. We have not included spin densities in Eq.~\ref{eq:energy-range-seperated} although a generalization of MC--srDFT with short-range spin density functionals is available.\cite{hedegaard2018b} In passsing, it should be noted that for an exact functional, Eq.~\eqref{eq:energy-range-seperated} is exact without the use of spin-densities. However, as realized for traditional DFT models, approximate functionals are generally significantly improved for open-shell systems by introducing spin-densities.

With the introduction of the range separation, the energy will in practice (for non-exact functionals) also depend on the range-separation parameter, $\mu$, that determines the extent of partitioning
between a long and short range descriptions: a value of $\mu=0$ corresponds to a pure DFT model, whereas a value of $\mu=\infty$ corresponds to a pure wave function model.  Based on analysis of small model systems a value of $\mu=0.4$ was previously suggested as this allocated the largest possible portion of dynamical correlation at the DFT functional.\cite{fromager2007}  This value was also optimal for excitation energies of organic systems\cite{hedegaard2013b,hubert2016a,hubert2016b,hedegaard2016}, but exploration of transition metal complexes have later shown that both $\mu=0.4$\cite{olsen2017,dong2018,jorgensen2022} and larger values\cite{kjellgren2021} can yield accurate results. 
\begin{figure*}[htb!]
	\centering
	\includegraphics[width=\textwidth]{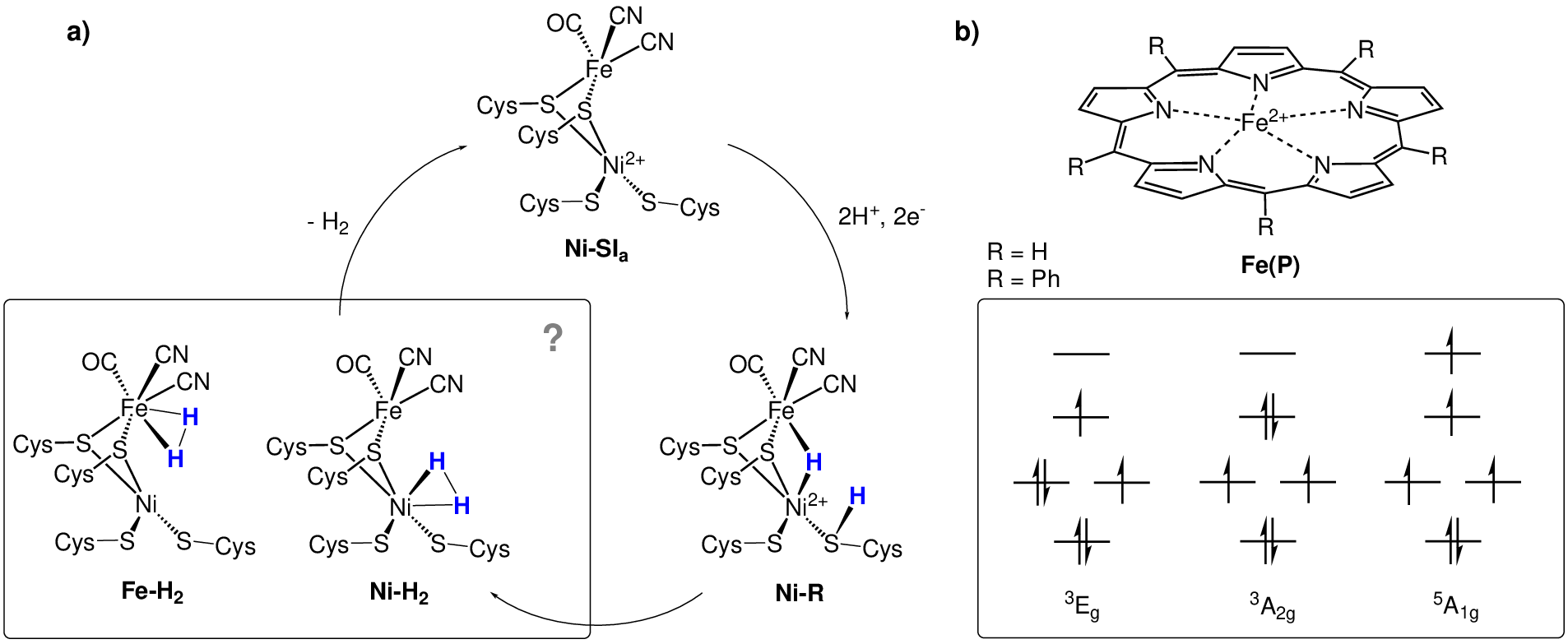}
	\caption{a) Selected intermediates in the reaction cycle of an [NiFe]-hydrogenase. Note that there is additionally two intermediates between \textbf{NI-Si}$_{\textbf{a}}$ and \textbf{Ni-R}, where the protons and electrons are added step-wise to \textbf{NI-Si}$_{\textbf{a}}$, initially to one of the cysteine ligands before one of the protons is moved to the bridging position on the \textbf{Ni-R} intermediate (see e.g.~Figure 2 of Ref.~\citenum{dong2017}).  b) Iron(II) porphyrin (with different R groups in the \textit{meso}-positions) with lowest three d$^6$ configurations and associated term symbol (assuming $D_{4h}$ symmetry).}
	\label{fig:NiFe-and-porph}
\end{figure*}

\section*{Beyond non-relativsitic Hamiltonians}

Before moving to selected case studies, we give a brief overview of the inclusion of relativistic effects within the framework of a multiconfigurational wave function. For a more in-depth account of the vast field of relativistic effects in quantum chemistry, we refer to the literature.\cite{dyall2007,reiher2014} We focus on explicit inclusion of relativistic effects, as opposed to the more implicit inclusion through effective core potentials (ECPs). The use of ECPs is firmly established and usually sufficient for transition metal active sites. We therefore focus on cases where this implicit inclusion is insufficient, namely L-edge spectroscopy which recently has begun to be employed for systems with relevance to bio-inorganic chemistry.\cite{wernet2019} The transitions observed in L-edge spectra involve electrons being excited from the metal 2s and 2p orbitals. Since the 2p orbitals are split due to relativistic effects, the 2p transitions are observed at two different energy ranges. These ranges are denoted "edges" with the symbols $\text{L}_2$, and $\text{L}_3$, respectively. 

The most direct method to include relativistic effects is to solve the  Dirac equation, which has the same form as the Schrödinger equation in Eq.~\eqref{eq:schrodinger}, but with a different form of the Hamiltonian, i.e., 
\begin{equation}
\hat{H} \rightarrow \hat{H}_D .
\end{equation}
We refer to the literature for the exact form of $\hat{H}_D $.\cite{dyall2007,reiher2014} Here it suffices to note that this new form will lead to a change of the wave function $\vert \Psi \rangle \rightarrow \vert \Psi_D \rangle$ to now having four components rather than just one as in the non-relativistic case. Methods that employ $\hat{H}_D$ directly with a multiconfigurational wave function do exist\cite{thyssen2008}, e.g., within the {\sc Dirac} program.\cite{saue2020} However, these methods have to the best of our knowledge not been employed within bio-inorganic chemistry, mainly due to their high computational cost. To reduce the cost, a common approach is to employ a unitary transformation of  $\hat{H}_D$ 
\begin{equation}
\hat{H}' = U^{\dagger} \hat{H}_D U ,
\label{eq:hamiltonian-relativistic}
\end{equation}
where $\hat{H}'$ has a much simpler form (this also leads to a simpler, two-component form of the wave function). Many different transformation matrices have been suggested in the literature (for a concise summary, see ref.~\cite{saue2011}). A commonly used form is the Douglas-Kroll-Hess transformation to second order (DKH2)\cite{douglas1974,hess1986}. The transformed Hamiltonian, $\hat{H}'$, can be divided into so-called scalar and spin-orbit coupling parts -- the latter are the ones crucial for L-edge spectroscopy. In the second-quantized form, the SO part is given 
\begin{equation}
\hat{H}'_{\text{SO}} =  \sum_{pq}\left( V^x_{pq}\hat{T}^x_{pq}  + V^y_{pq}\hat{T}^y_{pq}  + V^z_{pq}\hat{T}^y_{pq}   \right), 
\label{eq:so-operator}
\end{equation}
where $V^{\delta}_{pq}$ ($\delta = x,y,z$) are integrals over the spin-orbit coupling operator and $T^{\delta}_{pq}$ are triplet excitation operators, $T^x_{pq} = \frac{1}{2}(\hat{a}^{\dagger}_{p\alpha}\hat{a}_{q\beta} + \hat{a}^{\dagger}_{p\beta}\hat{a}_{q\alpha})$, $T^y_{pq} = \frac{1}{2}(-\hat{a}^{\dagger}_{p\alpha}\hat{a}_{q\beta} + \hat{a}^{\dagger}_{p\beta}\hat{a}_{q\alpha})$, and $T^z_{pq} = \frac{1}{2}(\hat{a}^{\dagger}_{p\alpha}\hat{a}_{q\alpha} - \hat{a}^{\dagger}_{p\beta}\hat{a}_{q\beta})$. As written, the SO operator in Eq.~\eqref{eq:so-operator} is an one-electron operator. Two-electron spin-orbit terms are also present, but it has been shown that these to a good approximation can be formulated as mean-field one-electron operators\cite{hess1996}.

The most common multiconfigurational method for calculating L-edge spectroscopy is of the RASSCF type;\cite{lundberg2019_Xray} the core (2p) orbitals are included in RAS1 (see Figure \ref{fig:CAS-RAS-fig}) and only states with a hole in this shell are generated, an approximation named core-valence separation.\cite{delcey19-hexs} The wave function is first optimized only including the scalar relativistic parts of $\hat{H}'$  (Eq.~\ref{eq:hamiltonian-relativistic}); hereby so-called spin-free states are obtained. These states are next allowed to interact through the SO operator in Eq.~\eqref{eq:so-operator} using the RAS state interaction approach (RASSI).\cite{malmqvist2002} It is the introduction of spin-orbit coupling that leads to the experimentally observed  splitting of the metal 2p orbitals (that otherwise are close to degenerate). When calculating these spectra, the transition energies are usually corrected with RASPT2. 

\section*{Case studies}

In this section, we briefly highlight recent studies of bio-inorganic systems with the different methods reviewed above. In all cases, we have chosen case studies where experimental work has been extremely complicated and different DFT methods have met problems: the first example is an enzyme with a small metal cluster, namely  the [NiFe]-hydrogenase where the active site contains Ni and Fe. For this case, both traditional CAS and approximate FCI solvers have been employed. Moreover, the MC--srDFT method has also been employed allowing detailed comparison between different methods. 

The next example shows that also metallo-enzymes with active sites containing a single metal atom can be problematic. The Fe(II) porphyrin is infamous for being a system where multiconfigurational wave functions and different DFT models obtain contradictory results. Correctly differentiating between different electron configurations would be expected to be a multiconfigurational problem. Yet, these methods were themselves struggling due to the restrictions imposed by the high computational cost of increasing the active space size. Only recently there has been some breakthroughs due to the large active space calculations made feasible by approximate FCI solvers.

\subsection*{[NiFe]-hydrogenase}

Hydrogenases are metallo-enzymes that catalyze the reversible conversion of \ce{H2} into protons and electrons -- a process that may become key for green hydrogen production and concomitant decarbonization of the energy sector. 
The hydrogenases are classified according to the architecture of the active site: The [Fe] and [FeFe]-hydrogenases contain a single or two Iron atoms, respectively, whereas the [NiFe]-hydrogenases contain a Nickel atom and an Iron atom in the active site.\cite{fontecillacamps2007} 

The [NiFe]-hydrogenases have been explored in numerous theoretical and experimental investigations.\cite{fontecillacamps2007,delacey2007,lubitz2007,lubitz2014,vincent2007,siegbahn2007} Based on these investigations, we show selected intermediates from the reaction cycle in Figure \ref{fig:NiFe-and-porph}(a). As seen from Figure \ref{fig:NiFe-and-porph}, the Iron atom is five-coordinated by the strong-field ligands \ce{CO} and \ce{CN-}, whereas two sulfur atoms from cysteine comprise the two remaining coordination sites. The two cysteine sulfur atoms form a bridge to the Nickel atom, which additionally is coordinated with two other (non-bridging) cysteine residues, leaving the Nickel atom four-coordinated.  Accordingly, both Iron and Nickel can receive an additional ligand (either \ce{H2} or \ce{H-}). 
\begin{table*}[htb!]
	\small
	\caption{Calculated energy differences between \textbf{Fe-H}$_{\textbf{2}}$ (either as $S = 1$ or $S = 0$) and \textbf{Ni-H}$_{\textbf{2}}$ using the \textbf{Ni-H}$_{\textbf{2}}$ with $S = 0$ as reference. Energies are in kcal /mol.\label{tab:NiFe}}
	\begin{tabular*}{1.00\textwidth}{@{\extracolsep{\fill}}llp{0.40\textwidth}lrcc}
		\hline \\[-1.5ex]
		Method                   & Basis & Model size (\# atoms)  & Spin &  $\Delta E$ & Binding site & Ref. \\[0.5ex]
		\hline\\[-1.0ex]
		B3LYP                    &  ECP/cc-pVDZ$^a$     &  NiFe cluster with \ce{\sbond S CH3} as Cys (30)                          & $S = 0$     & $-$6.6               &   Fe         &  \citenum{niu1999}\\[0.5ex]
		\hline \\[-1.5ex]
		B3LYP                    &  ECP/6-31G(d,p)$^a$  &   NiFe cluster with \ce{\sbond S CH2CH3} as Cys and Glu18, His72, Ser486 and Arg463    (94)$^d$    &    $S = 0$   &   8.9                & Ni              &  \citenum{wu2008} \\[0.5ex]
		B3LYP                    &  ECP/6-31G(d,p)$^a$  &   NiFe cluster with \ce{\sbond S CH2CH3} as Cys and Glu18, His72, Ser486 and Arg463    (94)$^d$    &    $S = 1$   & $-$1.2  & Fe                  &  \citenum{wu2008} \\[0.5ex]
		\hline \\[-1.5ex]
		TPSS                     &     def2-QZVP        &  NiFe cluster with \ce{\sbond S CH3} as Cys and Glu34 and His88 (47)$^e$                         & $S = 0$   &4.4      & Ni & \citenum{dong2017} \\[0.5ex]
		TPSS                     &     def2-QZVP        &  NiFe cluster with \ce{\sbond S CH3} as Cys and Glu34 and His88 (47)$^e$                         & $S = 1$   &  3.1    & Ni & \citenum{dong2017} \\[0.5ex]
		B3LYP                    &     def2-QZVP        &  NiFe cluster with \ce{\sbond S CH3} as Cys and Glu34 and His88 (47)$^e$                         & $S = 0$   & 4.1     & Ni & \citenum{dong2017} \\[0.5ex]
		B3LYP                    &     def2-QZVP        &  NiFe cluster with \ce{\sbond S CH3} as Cys and Glu34 and His88 (47)$^e$                         & $S = 1$   & $-$7.2  & Fe & \citenum{dong2017} \\[0.5ex]
		\hline \\[-1.5ex]
		QM/MM (TPSS)             &      def2-SV(P)     &   All residues within 4.5 \AA ~of the NiFe cluster and additional buried charges (819)            & $S = 0$    &  11.9   &  Ni   &  \citenum{dong2017} \\[0.5ex]
	    QM/MM (TPSS)             &      def2-SV(P)     &   All residues within 4.5 \AA ~of the NiFe cluster and additional buried charges  (819)           & $S = 1$    &  12.8   &  Ni   &  \citenum{dong2017} \\[0.5ex]
	    \hline \\[-1.5ex]     
		DMRG(22,22)-CASPT2       &     Mixed ANO$^b$   &  NiFe cluster with \ce{\sbond S CH3} as Cys  (30)                                            & $S = 0$  & 7.9  & Ni & \citenum{dong2017} \\[0.5ex]
		DMRG(22,22)-CASPT2       &     Mixed ANO$^b$   &  NiFe cluster with \ce{\sbond S CH3} as Cys (30)                                             & $S = 1$  & 7.7  & Ni & \citenum{dong2017} \\[0.5ex]
		DMRG(22,22)-CASPT2       &     Mixed ANO$^b$   &  NiFe cluster with \ce{\sbond S CH3} as Cys and Glu34 and His88 (47)$^e$                     & $S = 0$  & 2.7  & Ni & \citenum{dong2017} \\[0.5ex]
		DMRG(22,22)-CASPT2       &     Mixed ANO$^b$   &  NiFe cluster with \ce{\sbond S CH3} as Cys and Glu34 and His88 (47)$^e$                     & $S = 1$  & 6.2  & Ni & \citenum{dong2017} \\[0.5ex]
		CAS(10,10)-srPBE         &     cc-pVTZ/cc-pVDZ$^c$  &  NiFe cluster with \ce{\sbond S CH3} as Cys and Glu34 and His88 (47)$^e$                & $S = 0$  & 1.9  & Ni & \citenum{dong2018} \\[0.5ex]
		CAS(14,14)-srPBE         &     cc-pVTZ/cc-pVDZ$^c$  &  NiFe cluster with \ce{\sbond S CH3} as Cys and Glu34 and His88 (47)$^e$                & $S = 0$  & 2.5  & Ni & \citenum{dong2018} \\[0.5ex]
		\hline
	\end{tabular*}
	$^a$ The ECPs were used for both metals  and for heavier ligand atoms (see refs.~\citenum{niu1999} and \citenum{wu2008} for details). $^b$ Corresponds to VQZ on the metal, VTZ on the second-row atoms and DZ on hydrogen. $^c$ cc-pVTZ on Fe/Ni and cc-pVDZ on the remaining atoms.  $^d$ The residues numbers refer to the structure from \textit{D. gigas.}\cite{volbeda1995,volbeda1996}. $^e$ The residues numbers refer to the structure from \textit{D. vulgaris} Miyazaki F\cite{higuchi1999}
\end{table*}

We focus here on the intermediate where \ce{H2} binds or forms, depending on the direction of the reaction.  In this intermediate,  \ce{H2} is bound to either Fe (\textbf{Fe-H}$_{\textbf{2}}$) or Ni  (\textbf{Ni-H}$_{\textbf{2}}$). However, the preferred binding (or formation) site has been controversial: experimentally, \ce{CO} is a known inhibitor of \ce{H2} production and is known to bind to Ni\cite{ogata2002}. Further, gas-diffusion with Xe also leads to Ni as the binding site\cite{montet1997,volbeda2003}. Meanwhile, from a biological perspective, the coordination of \ce{CN-} and \ce{CO} ligands to Fe(II) is rather unique as it enforces a low-spin $d^{6}$ system on the Fe center. Low spin Fe(II) systems are known to bind \ce{H2} in organometallic chemistry\cite{bookh2ase1,bookh2ase2,kubas2007}, and this indicates that Fe is the \ce{H2} binding/formation site. 

The matter of the \ce{H2} binding/formation site has also been a target for theoretical investigations; a selection of results is provided in Table \ref{tab:NiFe}. Early investigations\cite{niu1999} with DFT predicted that \ce{H2} binds or forms at the Fe center (only the singlet state was investigated). In this early investigation, a small QM-cluster model was employed for the active site where the cysteine amino acids were represented by a \ce{\sbond SCH3} group.  Extending the QM-cluster size with four amino acids close to the active site (see Table \ref{tab:NiFe}) resulted in that \ce{H2}-binding to the Ni atom was more favorable by 8.9 kcal/mol\cite{wu2008} on the singlet potential energy surface (PES). Meanwhile, \ce{H2}-binding/formation on the triplet PES was also investigated, showing that this was 1.2 kcal/mol more stable than \ce{H2}-binding/formation at the Ni atom on the singlet PES. Thus, changing the size of the QM-cluster and/or the spin state may give qualitatively different results. This has previously been observed for investigations of the protonation state of cysteine residues in the [NiFe]-cluster\cite{hu2011,hu2013,sumner2013}  (the clusters in Table \ref{tab:NiFe} are therefore all selected to have the same protonation state). 

Another issue revealed by Table \ref{tab:NiFe} is that different functionals give different conclusions regarding the most stable site (and spin) for \ce{H2} binding/formation: systematic investigations by Ryde and co-workers\cite{dong2017} showed for a system comprised of the [NiFe] cluster and the two closest amino acid residues (Glu34 and His88 in the investigated hydrogenase), the hybrid GGA B3LYP and the pure \textit{meta}-GGA functional TPSS both find that \textbf{Ni-H}$_{\textbf{2}}$ is most stable on the singlet PES. Yet, the hybrid functional predicts that the \textbf{Fe-H}$_{\textbf{2}}$ intermediate is more stable on the triplet PES, unlike the TPSS functional. Thus, DFT is not sufficiently accurate for a conclusive result.   

For a long time, calculations with multiconfigurational methods were not possible for the [NiFe]-hydrogenase active site due to the large requirement for the active space imposed by the two metal atoms. This has now changed with the approximate FCI solvers discussed above. In a paper by Dong \textit{et al.}\cite{dong2017}, CASPT2 calculations based on a large CASSCF(22,22)  were carried out using DMRG as the approximate FCI method. Selected results are given in Table \ref{tab:NiFe}, denoted DMRG(22,22)-CASPT2. They all show that the singlet \textbf{Ni-H}$_{\textbf{2}}$ is the most stable. Unfortunately, current implementations of DMRG cannot treat a large number of inactive orbitals, and hence larger [NiFe]-clusters cannot be treated. Instead, a DFT calculation was performed in which all residues within 4.5 {\AA} as well as all buried charges were included, finding that this further stabilized the  \textbf{Ni-H}$_{\textbf{2}}$ state.\cite{dong2017} We expect that the developments in programs like {\sc MultiPsi} discussed above will make it feasible to avoid resorting to DFT to perform these types of calculations.
% Cite RASPT2 of delcey2014?

The MC-srDFT method has been employed for the same hydrogenase model  as used by Dong \textit{et al.}\cite{dong2017}, so direct comparison to CASPT2 with large active spaces from DMRG is possible: We denote the method CAS-srPBE in Table \ref{tab:NiFe}. The CAS-srPBE method also predicts that the \textbf{Ni-H}$_{\textbf{2}}$ state is most stable\cite{dong2018}.  Interestingly,  CAS(14,14)-srPBE obtain a result close to the DMRG-based method, despite using a much smaller active space:  CAS(14,14)-srPBE obtains an energy difference of 2.5 kcal/mol versus 2.1 kcal/mol with DMRG(22,22)-CASPT2. Even reducing to a CAS(10,10) does not change the results dramatically. The large difference in active spaces is primarily because no double shell of 3d orbitals was included in the CAS-srPBE calculations, making the active space requirement considerably smaller. At the time Ref.~\citenum{dong2018} was published, no open-shell implementation of the MC-srDFT method existed\cite{hedegaard2018b}  so the triplet state could not be obtained.

Thus, the calculations on [NiFe]-hydrogenase illustrate that multiconfigurational calculations with large active spaces and larger cluster sizes have become feasible. Using this system as a benchmark, it also shows that multiconfigurational hybrids with DFT show promising results in recovering dynamical correlation efficiently. 

%The most stable spin-state  of the  Ni-SI$_{\text{a}}$ intermediate has long been debated. Magnetic spectroscopies and saturation measurements have suggested a low-spin d$^8$ configuration\cite{kowal1988,dole1997,wang1992}, whereas Ni L-edge X-ray absorption spectroscopy is commensurate with a high-spin state.\cite{wang2000} Unfortunately, theoretical investigations based on DFT cannot reach a conclusion, since different hybrid functionals such as B3LYP have favored the high-spin state and non-hybrid functionals such as BP86 favor low-spin states.\cite{fan2001,Bruschi2004,pardo2006,jayapal2008,yson2013,visser2014}    

\subsection*{Iron-porphyrin}

The Iron porphyrin complex shown in Figure \ref{fig:NiFe-and-porph}(b) is a model system of the haem group which  assumes a key position for aerobic life. Some proteins use the haem group to either transport or store \ce{O2} (haemoglobin and myoglobin). Many other proteins instead use the haem-group to activate \ce{O2} to participate in chemical transformations .\cite{poulos2014,huang2018,bookcytochrome} An example of such activation is the cytochrome P450 enzymes which oxidize otherwise inactive \ce{C-H} bonds. The haem group is also found in peroxidases, such as cytochrome \textit{c} peroxidase and horseradish peroxidase.\cite{shaik2010,huang2018} These enzymes protect against oxidative stress in cells by removal of hydrogen peroxide.

The resting state of the iron-haem proteins contains d$^{6}$ Fe(II). However, the electronic ground state of the haem model system Fe(II) porphyrin is still not assigned. The ligand field imposed by the porphyrin ligand gives rise to several possible electronic states. The ligand-field orbital splittings of the three lowest of these states, $^3$E$_{g}$, $^5$A$_{1g}$, and $^3$A$_{2g}$, are shown in Figure \ref{fig:NiFe-and-porph} (b) along with the associated term symbol (assuming $D_{4h}$ symmetry). Different spectroscopic techniques have suggested the $^{3}$A$_{2g}$ state to be lowest\cite{collman1975,lang1978,boyd1979,goff1977,mispelter1980}, but resonance Raman spectroscopy has been used to argue for $^{3}$E$_{g}$ instead.\cite{kitagawa1979} L-edge spectroscopy has also been applied\cite{wilson2013}, but the spectra are difficult to intepret without theoretical support. Comparisons between the experimental studies are further complicated by the different variants of the porphyrin skeleton (i.e.~different \ce{R} groups, cf.~Figure \ref{fig:NiFe-and-porph}b) use in  the experimental investigations.
 \begin{table}[t]
	\small
	\caption{Calculated spin-state splittings of porphyrin with R = H, see Figure \ref{fig:NiFe-and-porph}(b),  using different methods. Energies are in kcal /mol.}
	\label{tbl:feporph}
	\begin{tabular*}{0.48\textwidth}{@{\extracolsep{\fill}}llrl}
		\hline \\[-1.5ex]
		Method & Basis &   Splitting & Ref. \\[0.5ex]
		\hline\\[-1.0ex]
		\multicolumn{4}{c}{ $^{3}$E$_{g}$ $-$ $^{5}$A$_{1g}$}   \\[0.5ex] 
		B3LYP                                                   & VTZ (Fe)/6-31G       & $-0.7$   & \citenum{kozlowski1998}\\[1.0ex]
		CASPT2(14,16)                                     & ANO-VTZP                  & $0.5$   & \citenum{limanni2018}\\
		CASPT2(16,15)                                      & ANO-VTZP/VDZP   &  $1.4$ & \citenum{vancoillie2011}   \\[1.0ex]
		RASSCF(32,34)$^a$                          & ANO-VTZP            & $2.9$     & \citenum{limanni2018}\\
		DMRG-CI(32,34)$^d$                       &  ANO-VTZP              &   $-3.1$ & \citenum{weser2021}   \\ 
		Stochastic CASSCF(32,34)$^b$   & ANO-VTZP               & $-3.1$       & \citenum{limanni2018}\\
		Stochastic CASSCF(40,38)$^c$    & ANO-VTZP              & $-4.4$ & \citenum{limanni2019} \\
		CCSDTQ-CAS(40,38)                         & ANO-VTZP               & $-4.8$  & \citenum{limanni2019} \\[1.0ex]
		CAS(16,15)-fPBE                                  & Mixed ANO$^f$    & $-12.9$  & \citenum{zhou2019} \\
		DMRG(34,35)-fPBE$^e$                           & Mixed ANO$^f$   & $-16.6$  & \citenum{zhou2019} \\
		\\[1.0ex]
		\multicolumn{4}{c}{ $^{3}$A$_{2g}$ $-$ $^{5}$A$_{1g}$ }\\[0.5ex]
		B3LYP                                                   & VTZ (Fe)/6-31G     & $-6.9$      & \citenum{kozlowski1998}\\[1.0ex]    
		CASPT2(16,15)                                     & ANO-VTZP/VDZP  &  $-0.9$   & \citenum{vancoillie2011}   \\[1.0ex]
		RASSCF(32,34)$^a$                          & ANO-VTZP              & $4.1$     & \citenum{limanni2018} \\[1.0ex]
		Stochastic CASSCF(32,34)$^b$ & ANO-VTZP              & $-2.6$  & \citenum{limanni2018} \\[1.0ex]
		CAS(16,15)-fPBE                                  & Mixed ANO$^f$   & $-12.9$   & \citenum{zhou2019} \\
		DMRG(34,35)-fPBE$^e$                            & Mixed ANO$^f$   & $-15.7$  & \citenum{zhou2019} \\
		\hline
	\end{tabular*}
	$^a$ Two holes in RAS1 and two electrons in RAS3 allowed. $^b$ 1B walkers. $^c$ 4B walkers. $^d$ $m = 10000$. $^e$ $m = 300$ $^f$ Corresponds to VQZ on the metal, VTZ on the second-row atoms and DZ on hydrogen.
	\label{tab:feporph}
\end{table}
\begin{figure}[htb!]
	\centering
	\includegraphics[width=0.48\textwidth]{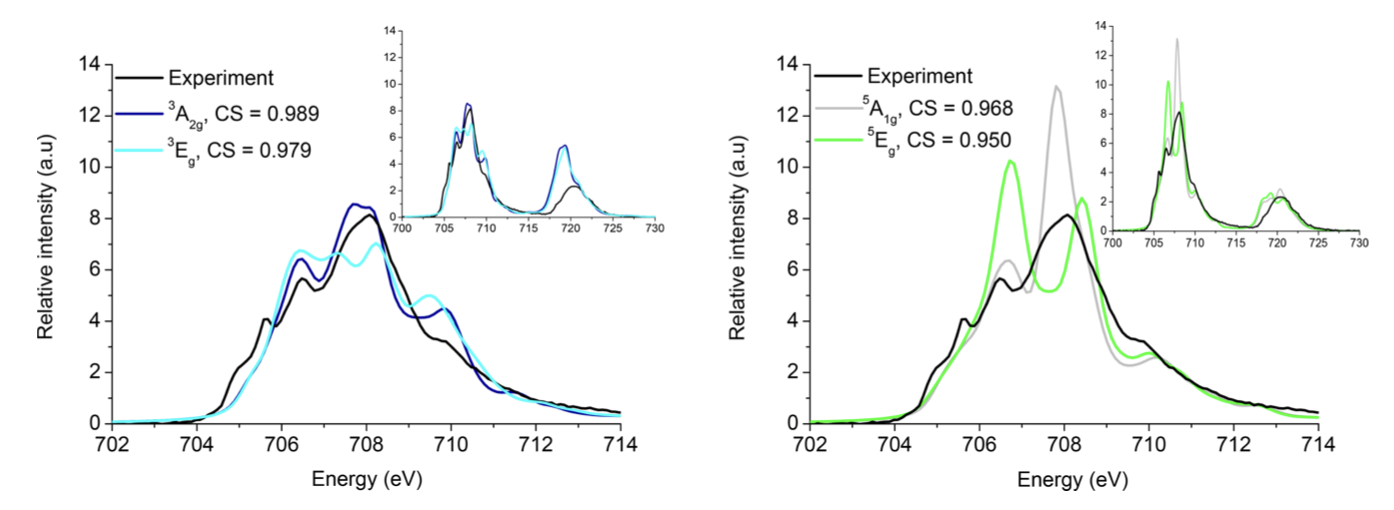}
	\caption{L$_3$-edge X-ray absorption spectra  details of the different possible ground state electronic structure of the iron-porphyrin computed with RASPT2 compared to the experimental spectrum. CS indicates the cosine similarity between the experiment and theory. Figures extracted from ref. \citenum{guo19-hemexray}.}
	\label{fig:heme-ledge}
\end{figure}

Theoretical studies have also attempted to clarify what the lowest electronic state of Fe(II) porphyrin is. We focus here on the simplest Fe(II) porphyrin (\ce{R = H}) where most high-level calculations exist. We have compiled selected results for the relative differences between the lowest $^3$E$_{g}$, $^5$A$_{1g}$, and $^3$A$_{2g}$ states in Table \ref{tab:feporph}. Many DFT functionals obtain a triplet ground state\cite{chen2011} --we use here B3LYP in Table \ref{tab:feporph} to represent DFT; B3LYP predicts $^3$A$_{2g}$ to be the lowest state.
Meanwhile, employing CASPT(14,16) leads to $^{5}$A$_{1g}$ being the ground-state.\cite{manni2014} Notably, this CAS corresponds to including the d-orbitals along with an extra set of ("double shell") d-orbitals, the Fe 4s orbital, and selected orbitals in the ligand macrocycle in the active space. In another investigation, it was found that a CAS(16,15) that included the Iron 3s/3p orbitals yielded the $^3$A$_{2g}$ state 2.3 kJ/mol below the $^{5}$A$_{1g}$ state while the $^3$E$_{g}$ was 1.4 kcal/mol above $^{5}$A$_{1g}$.  The effect of increasing the active space could for a long time not be investigated with traditional CAS solvers. However, with the approximate FCI solvers based on FCI-QMC\cite{limanni2016}, a stochastic CASSCF(32,34) \cite{limanni2016,limanni2018} could be employed, leading to the triplet states being below the $^{5}$A$_{1g}$ state by $3.1$ kcal/mol ($^{3}$E$_{g}$ ) and $2.6$ kcal/mol ($^{3}$A$_{2g}$). Thus, the  $^{3}$E$_{g}$ state  becomes the ground state. A similar result was obtained with DMRG.\cite{weser2021}, while RAS-type restrictions of the CAS(32,34) active space still leads to $^{5}$A$_{1g}$  being the ground state (cf.~Table \ref{tab:feporph}). 

Since semi-core correlation has been shown to be important for energy differences between different spin states\cite{pierloot17-3s3p}, a later paper investigated how extension of the CASSCF(32,34) with the four Iron 3s/3p orbitals affects the  difference between the $^{5}$A$_{1g}$ and $^{3}$E$_{g}$  states.\cite{limanni2019} The resulting stochastic CASSCF(40,38) still yield $^{3}$E$_{g}$ lower than  $^{5}$A$_{1g}$, now by $4.1$ kcal/mol. A coupled-cluster calculation including up to quadruple excitations within the same CAS(40,38) gives a similar result\cite{limanni2019}, as evident from Table \ref{tab:feporph}.

Interestingly, a recent investigation with pair-density functionals using both traditional CAS solvers with a CAS(16,15) active space, and large CAS(34,35) active space from approximate FCI solvers, also obtained the triplet states below the $^{5}$A$_{1g}$ state\cite{zhou2020}, similar to the pure wave function methods discussed above. The size of the splitting is somewhat larger (cf.~Table \ref{tab:feporph}), but whether this is a functional artifact or insufficient recovery of the dynamical correlation in the pure wave function based methods remains to be seen.   

Multiconfigurational calculations have also been used to simulate the iron L-edge X-ray absorption spectrum, using comparison with the experimental spectrum.\cite{wilson2013,guo19-hemexray,lundberg2019_Xray} as an alternative route to identify the electronic structure of the ground state (see Figure \ref{fig:heme-ledge}). Even with a modest CAS(8,11) active space for the valence electrons, the triplet $^{3}$A$_{2g}$ state was found to have the best agreement with the experimental spectrum, though the agreement could be improved further by introducing a fraction of quintet through spin-orbit coupling. 

Finally, it should not be forgotten that most experimental work employ substituted porphyrins, e.g., with R = Ph. Moreover, if a water-soluble porphyrin skeleton is desired, even large groups such as sulfonatophenyl are required.\cite{mazzeo2019}  While sulfonatophenyl groups are not very likely to affect the choice of active space, all results in Table \ref{tab:feporph} have mainly focused on the simplest form to avoid a large number of inactive orbitals. However, recently results with the {\sc MultiPsi} program have showed that calculations with large sulfonatophenyl groups are now within reach, also for calcuclating excited states and spectroscopy.\cite{scott23-cpp}

\section*{Conclusions}

We have given an overview of recent progress in multiconfigurational wave functions, focusing on developments promising for bio-inorganic chemistry. One of the key developments over the last years is the push for increasingly large active spaces through approximate FCI methods. We explicitly discussed the DMRG and FCI-QMC methods that can provide results close to FCI for large active spaces at a fraction of the cost. Within the last years, FCI-QMC and DMRG have enabled applications of multioconfigurational wave functions on large metal-cluster active sites. We illustrate these methods on a somewhat smaller bi-metallic active site, namely the active site of [NiFe]-hydrogenase. In this case, the approximate FCI solvers were employed combined with perturbation theory to recover dynamical correlation; a calculation that few years ago would have been impossible. We also discussed the advances these methods have brought to understanding the electronic structure of haem groups - a case where multiconfigurational methods for a long time obtained a ground state that could not be reconciled with experiment. Recent results with FCI-QMC and DMRG show that the employed active space sizes in earlier calculations with traditional CASSCF solvers presumably were too small. 

While large active spaces are important for many types of protein active sites, no approximate FCI method can recover dynamical correlation entirely through the active space. We have briefly covered traditional perturbation-theory-based methods for this task. %We illustrate a recent application of these methods to a recently discovered metalloenzyme family, namely the LPMO enzymes; an enzyme family that seemingly performs the same chemistry as some haem groups, but whose mechanism is yet to be unraveled. Attempts with DFT have shown that this method meets issues for several of the important intermediates.  
We  compare perturbation-theory-based methods to new models that avoid perturbation theory and instead combine a multiconfigurational wave function with DFT functionals. While these methods are still in development, a great advantage is that they have lower scaling and avoid higher-order reduced density matrices. The computational cost is therefore on-par with that of performing a CAS(SCF). The first applications on [NiFe]-hydrogenase and Fe porphyrin are promising. 

Finally, we have pointed out that bio-inorganic systems often require that we can handle more than 100 atoms. Thus,   CAS(SCF) solvers that can handle many inactive electrons (i.e.
~many basis functions) will become indispensable in the future. Development of efficient CAS(SCF) solvers has only very recently received focus. As illustrated in this perspective, the {\sc MultiPsi} program can handle a whole metalloprotein with both CASSCF and different CAS-DFT hybrid models. Thus, we are now entering a new age for bio-inorganic systems where multiconfigurational methods can be employed beyond small models.  We expect this will lead to new insights in a field that so far has been dominated by single-reference methods.  

\section*{Author Contributions}

All authors contributed equally to the writing of the manuscript. 

\section*{Conflicts of interest}
There are no conflicts to declare.

\section*{Acknowledgements}
We acknowledge the The Villum Foundation, Young Investigator Program (grant no.~29412), the Swedish Research Council (grant no.~2019-04205), and Independent Research Fund Denmark (grant no.~2064-00002B) for support. 

%%%END OF MAIN TEXT%%%


\balance


%%%REFERENCES%%%
\bibliography{bib} %You need to replace "rsc" on this line with the name of your .bib file
\bibliographystyle{rsc} %the RSC's .bst file

\end{document}